# Initial energy deposition and initiation mechanism of nanosecond laser damage caused by KDP surface micro-defects

Hao Yang[1,2], Mingjun Chen[1], Jian Cheng[1,*], Zhichao Liu[2], Linjie Zhao[1], Qi Liu[1], Chao Tan[1], Jian Wang[2]

[1]State Key Laboratory of Robotics and System, Harbin Institute of Technology, Harbin 150001, China

[2]Research Center of Laser Fusion, China Academy of Engineering Physics, Mianyang 621900, China

Correspondence:
Dr. Jian Cheng, State Key Laboratory of Robotics and System, School of Mechatronics Engineering, Harbin Institute of Technology, Harbin 150001, China.
Tel.: +86(0)451-86403252. Fax: +86(0)451-86403252.
*E-mail: cheng.826@hit.edu.cn*

**Abstract:**

To enable an exploration of the initiation mechanism of nanosecond laser damage on a potassium dihydrogen phosphate (KDP) surface, a defect-assisted energy deposition model is developed that involves light intensity enhancement and a sub-band gap energy level structure. The simulations provide an explanation on why the laser-induced damage threshold (LIDT) of the KDP crystal is two orders of magnitude lower than the theoretical value. The model is verified by use of the transient images that appear during the laser damage. In addition, the dimensions of the "dangerous" surface defects that are the most sensitive to the laser damage are proposed. This work enables clarification on the initial energy deposition (IED) and initiation mechanism of the nanosecond laser damage caused by the KDP surface defects on micro-nano scale. It is helpful in understanding the laser-matter interactions and to improve the processing technique for high quality optical components.

# 1. Introduction

Potassium dihydrogen phosphate (KDP) crystals are typically employed as photoelectric switches and frequency converters in high-power laser facilities [[1]-[3]]. These types of crystals crucially contain a wide energy band gap (~ 7.7 eV) and a low linear absorption coefficient (~ 0.005 cm$^{-1}$). Therefore, the intrinsic laser-induced damage threshold (LIDT) of a perfect KDP crystal can theoretically match the order of TW/cm$^2$ magnitude [[4]]. However, practical KDP crystals processed by the single-point diamond fly-cutting inevitably undergo laser damage when irradiated by a nanosecond laser with a magnitude of GW/cm$^2$ [[5]]. Therefore, the issues relating to laser damage to KDP crystals seriously limits the development of high-power laser facilities at present.

In the last five decades, the laser damage of optical components has become a hot research topic. For example, Bloembergen *et al.* [[6]] proposed the electron avalanche breakdown model to analyze such damage. They believe that, during laser irradiation, free electrons produce more free electrons via collision ionization of the other atoms in the medium. The laser damage occurs when the amount of local free electrons reaches a certain limit. Based on a multi-wavelength laser damage test on bulk KDP crystals, Carr *et al.* [[7]] determined that the sub-band gap energy level structure of the material changes when close to the lattice defects inside the KDP crystals. The defect-assisted multi-photon absorption mechanism is used to explain the bulk damage initiation phenomenon, when the laser single-photon energy is lower than the band gap of the KDP material. Moreover, surface damage to the optics is generally considered to relate to the surface defects that are introduced by the processing. Cheng *et al*. [[8]] found that the geometric morphology of the surface defects caused a light field modulation and predicted that the laser damage resistance of the KDP crystal depends on the local light intensity enhancement near to the surface defect. Wang *et al*. [[9]] built the equivalent 'explosion' model to study the fracture process of the KDP crystal during laser damage. Based on the multi-physics field coupling algorithm, Yang *et al*. [[10]] solved the hydrodynamics behavior of the laser damage caused by the surface defects on the KDP crystals and verified the influence of the phase change of the material and the fluid flow on the laser damage evolution. However, the current research is mostly focused upon the boule material [[6],[7]] or the laser damage formation process [[8]]-[10]]. Studies on the initial energy deposition (IED) mechanism, which relates to the laser damage caused by surface defects on the KDP crystals, are lacking. Although the light intensity enhancements that are induced by the surface defects can partly explain the mechanism of the laser damage, the magnitude of the light intensity enhancement is not sufficient to reach the intrinsic threshold of the optics [[11]]. The IED mechanism and the damage initiated during the low-power-density nanosecond laser irradiation on the KDP crystal surface remains unclear.

In this work, we establish the defect-assisted IED model by combining the geometric morphology characteristics with the material sub-band gap energy level structure of the KDP surface defects. Based on the results of the model, the laser energy deposition near the surface defect of the KDP crystals, under low power density nanosecond laser irradiation, are quantificationally analyzed to reveal the initiation mechanism. Combining the theory of the defect-assisted light intensity enhancement and the material ionization, the initiation process of laser damage is explained more comprehensively, and the understanding of the laser damage mechanism becomes more developed. The model also successfully provides the reason why the surface damage of the KDP crystals is far less than the theoretical LIDT. This discovery enables suggestions on the direction for improving the laser damage resistance of the KDP crystals, thus optimizing the processing technology in future.

## 2. Theory and Experiment

The indentation and the scratch defects are first prepared on the surface of the KDP crystals (Fig.1). There are many lateral cracks around the indentation and the scratch [[12]]. The lateral cracks are caused by a sideways extrusion of the material. The cracks extend laterally along the element surface. Figures 1(a) and 1(b) show the fluorescence images of the surface indentation and the scratch, respectively. The fluorescence intensity of the positions near to the lateral crack is much higher than those of the indentation and the scratch itself. The fluorescence intensity is strongly related to the laser absorption characteristics of the material [[13]]. The stronger the fluorescence intensity, the more intense the laser energy absorption. The material close to the lateral crack has a strong absorption effect on the laser. The two-dimensional cross-sectional profile of the crack around the indentation and the scratch is shown in Fig. 1(c). The heave height at the crack can reach ~ 240 nm and ~ 380 nm. The opening width of the crack is much smaller than the overall crack width, generally only ~ 1 μm. Figure 1(d) shows the scanning electron micrograph of the crack surface on the KDP crystals. The crack has an arc-shaped surface inside, and the crack surface close to the element surface is rough. This indicates that the crack starts on the crystal surface and it then extends into the crystal interior. In the fluorescence image, the fluorescence intensity is at its strongest at these locations. The material near to the crack is modified and the sub-band gap energy level structure appears, which is straight-forward to generate for a local electron excitation under laser irradiation [[13],[14]].

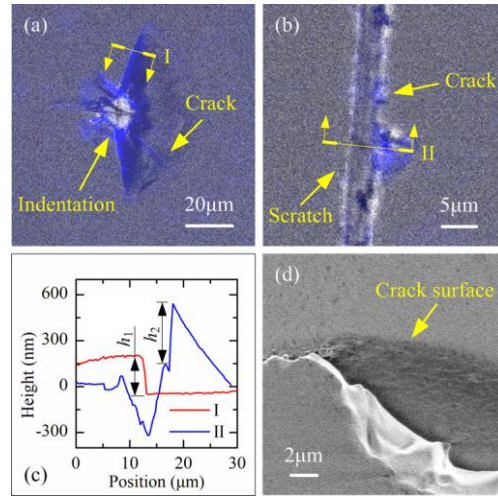

Fig. 1 Micromorphology of the surface defects on the KDP crystals. (a) Fluorescence image of the indentation defects with the surrounding cracks on the KDP crystals. (b) Fluorescence image of the scratch defect with surrounding cracks on the KDP crystals. (c) 2D cross-sectional profile of the cracks around the indentation and the scratch defects on the KDP crystals surface. (d) Scanning electron micrograph of crack surface on the KDP crystals.

Considering the geometric morphology of the lateral crack on the KDP crystals surface, and the change in the sub-band gap energy level structure, the study of an IED model (coupled with light intensity enhancement and the energy level structure) is now proposed to analyze the energy deposition mechanism near the surface defects on the KDP crystals under a nanosecond laser irradiation. The special geometric structure of the surface defects causes the laser light field modulation, which results in local hot spots of light intensity [[8]]. During the interaction with the laser pulse, the internal electronic structure of the material close to the defects contains a series of excited states [[7],[14]]; the energy gap between the first and the second excited states can be bridged by $3\omega$ single-photon absorptions. Valence-electron-excitations band transitions are promoted by the sequential single photon absorptions. The IED is calculated by combining the electronic attenuation and the inverse bremsstrahlung regime.

This IED model omits consideration of the internal energy diffusion of the material. The subsequent energy absorption, and transport under laser irradiation, can be solved by using the previously established dynamics model that couples with the multi-physical field [[10]]. The lattice size of the KDP crystals is ~ 10 nm. Based on a simple calculation, the energy transfer between the KDP crystal lattices starts after ~ 40 ps. Therefore, the time of 40 ps is used as the time boundary between the IED model and the subsequent dynamics coupling model of laser damage. The IED results are used as the prerequisite for the subsequent calculation of the dynamics model. Finally, the material phase transition and the fluid flow during the laser damage of the KDP crystals are initiated.

## 3. Results and discussions

Figure 2 shows the calculations of the IED model caused by a lateral crack on the KDP surface under laser irradiation at 40 ps. In the model, the lateral crack width is assumed to be 600 nm, the laser wavelength is 355 nm, and the power density is 5 GW/cm$^2$. The local electron density distribution close to the lateral crack at 40 ps is shown in Fig. 2(a). The electron density in the KDP crystals is generally at a low level. Moreover, there is a small amount of electron accumulation towards and at the bottom of the lateral crack. The electron density there is no greater than $1 \times 10^{21}$ cm$^{-3}$, which has little effect on the laser energy deposition. However, due to the combined effects of light field modulation and the sub-band gap energy level structure, the free electron density increases dramatically, and eventually exceeds $3 \times 10^{21}$ cm$^{-3}$ at the lateral crack center. The increase of the free electron density is extremely helpful for the laser energy deposition. Figure 2(b) shows the local temperature distribution near to the lateral crack on the KDP crystals surface at 40 ps. A high temperature region with a local temperature ~ 4931 K appears at the center of the latera crack. The ellipsoid high-temperature region is the IED center; here, it is extremely easy to induce the laser damage initiation under the subsequent laser irradiation.

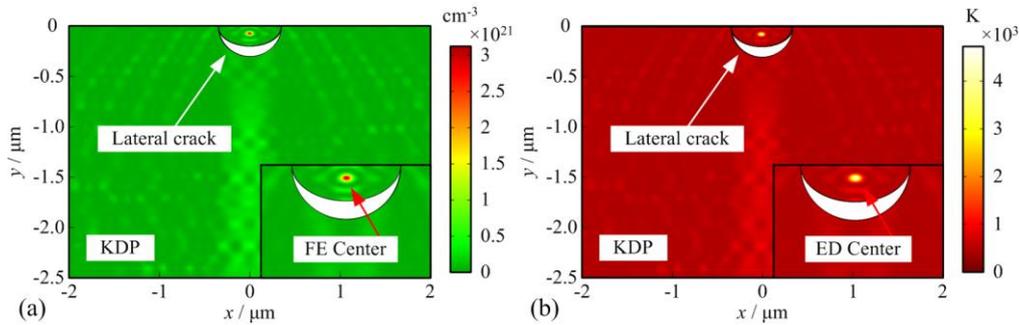

Fig. 2 Graphs of the calculations of the IED model caused by the lateral crack on the KDP surface under laser irradiation at 40ps. (a) Local electronic density distribution near to the lateral crack on the KDP crystals surface at 40ps. (b) Local temperature distribution in the vicinity of the lateral crack on the KDP crystals surface at 40ps.

By plotting graphs using the IED model of the lateral cracks with various widths, the curves of the electronic density and the temperature against the widths of the lateral cracks on the KDP crystals surface are shown in Fig. 3. The two curves show the same trends. With an increase in the lateral crack width, the values of both curves generally increase. However, the curves are seen to decrease at the width of 800 nm for a period. When the width is greater than 1200 nm, the curves again have an upward trend. From the results of the dynamics model coupled with the multi-physical field of the laser damage, it is inferred that (when the local temperature caused by the IED is greater than 5000 K) the laser damage on the KDP crystals occurs under subsequent laser irradiation. According to the critical temperature criterion, the lateral cracks with widths that range from 600 nm to 1100 nm and greater than 1300 nm will cause laser damage at 40 ps. However, based on the critical electron density criterion (8.85 $\times 10^{21}$ cm$^{-3}$), only the lateral crack with a width greater than 1400 nm can induce the avalanche ionization at 40 ps. This indicates that the laser damage caused by the surface defects on the KDP crystals surface at low energy density is not caused by the electrical breakdown. For example, for defects with a width of 600 nm, the local temperature (at 40 ps) extends into the critical values and the laser damage is initiated. However, the electronic density did not reach the critical breakdown density at this time. Thus, the temperature rise is the critical factor to trigger the laser damage. This differs from some previous studies that analyze the breakdown regime [[6]]. However, research on the strongly nonlinear laser energy absorption process supports our viewpoint [[15]]. With the increase of the subsequent laser irradiation time, the material phase transition and the hydrodynamics go into effect after the local temperature reaches a critical value. This temperature rise, which is caused by the IED, is the key factor for the initiation of the laser damage. The local maximum temperature inside the KDP crystals can be used as the laser damage criterion. The lateral cracks, with the widths ranging from 600 nm to 1100 nm and greater than 1300 nm, are vulnerable under laser irradiation; this should receive much more attention in relation to processing and application.

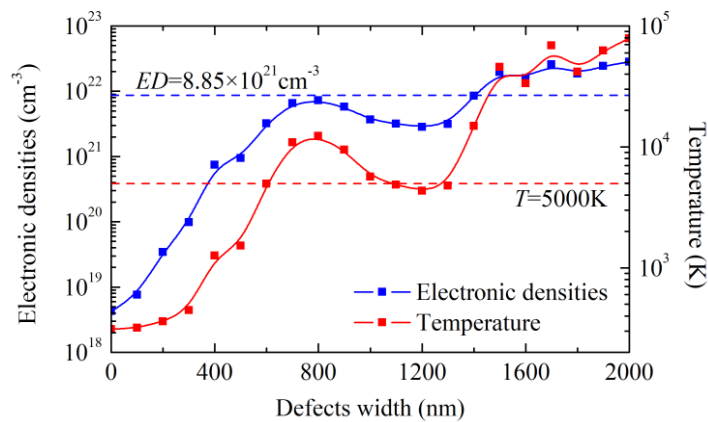

Fig. 3 Graph showing the curves of the electronic density and temperature against the widths of lateral cracks on the KDP crystals surface, calculated by the IED model.

To verify the defect-assisted IED model, and the initiation mechanism of the laser damage, we design a time-resolved system [[17],[18]] to capture the transient images of the laser damage that is caused by the lateral cracks on the KDP crystals. Figure 4 shows the resultant images. Here, there are black shadows in the positions near to the lateral cracks, when the laser is irradiated for 1-2 ns. These black shadows are morphologically similar to the simulated high-temperature region in Fig. 2(b). This is due to the strong absorption of laser light by ionized materials. On comparing the final microscopic images of the indentation and the scratch, it is determined that the position of the IED strictly corresponds to the final damage sites. The results demonstrate the validity of the defect-assisted IED model for the KDP crystals. Besides, it reveals that the IED caused by surface defect, under nanosecond laser irradiation, plays a vital role in the laser damage initiation.

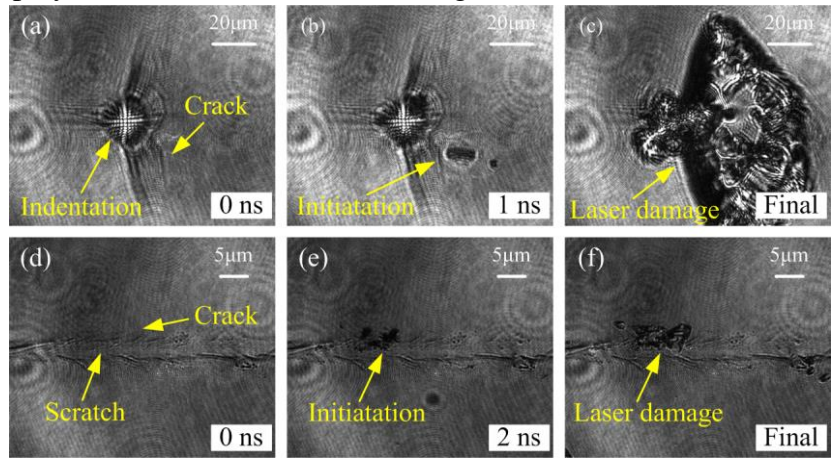

Fig. 4 Transient images of the laser damage, caused by the lateral cracks on the KDP crystals, that are taken by the time-resolved system. (a-c) Microscopic images of the lateral cracks around the indentations on the KDP crystals surface after the laser irradiation at 0 ns, 1 ns and the resulting final damage site. (d-f) Microscopic images of lateral cracks around the scratch on the KDP crystal surface after laser irradiation at 0 ns, 2 ns and the resulting final damage site.

## 4. Conclusion

In summary, the mechanism of the laser damage that is induced by surface defects on the KDP crystals, at low laser power density, has been revealed. The surface defects of the KDP crystals have special geometric morphology and a sub-band gap energy level structure. Based on the theory of laser propagation and electron excitation, an IED model coupling to the light intensity enhancement and the energy level structure is established. This solves the evolution of the free electron density and the temperature at the initial stage of laser irradiation. The surface defect morphology causes the light field modulation, producing local hot spots of light intensity. The sub-band structures of the material close to the surface defects contributes to the sequential one-photon absorptions relating to $3\omega$. The local temperature rise, caused by the electron non-radiative relaxation, is the dominant factor for the damage initiation. The material temperature can, thus, be used as a laser damage criterion. The IED plays a triggering role in the subsequent heat conduction and fluid flow. The IED model, and the initiation mechanism of the defect induced laser damage, were also proved by laser damage experiments. Moreover, we have shown that lateral cracks with a width range from 600 nm to 1100 nm or ≥1300 nm create vulnerabilities when irradiated by ns lasers. This model successfully provides a reason for why the surface damage of the KDP crystals is far less than the theoretical LIDT. It also offers a direction for improving the laser damage resistance, which will aid in the development of KDP crystals processing technology in the future.

**Funding**

National Natural Science Foundation of China (Nos. 51775147, 51705105); Science Challenge Project (No. TZ2016006-0503-01); Young Elite Scientists Sponsorship